\documentclass[conference]{IEEEtran}

\usepackage[cmex10]{amsmath}
\usepackage[cmex10]{amsmath}
\usepackage{array}
\usepackage{verbatim}
\usepackage{tikz}
\usepackage{epstopdf}
\usepackage{amsfonts}
\usepackage{amsmath}
\usepackage{amssymb}
\newtheorem{defi}{Definition}
\newtheorem{theorem}{Theorem}
\newtheorem{lemma}{Lemma}
\newtheorem{remark}{Remark}
\newtheorem{corollary}{Corollary}

\newcommand{\sm}{\sum _{\ell=1}^K}
\newcommand{\modn}{\hspace{-0.1in}\mod\Lambda}
\newcommand{\modnjl}{\hspace{-0.1in}\mod\Lambda_{(j,\ell)}}
\newcommand{\modnli}{\hspace{-0.1in}\mod\Lambda_{(\ell,i)}}
\newcommand{\modnll}{\hspace{-0.1in}\mod\Lambda_{(\ell,\ell)}}

\usepackage{cite}
\usepackage[center]{caption}
\usepackage{graphicx}
\usepackage{multicol}
\begin{document}
\title{Finite-SNR Regime Analysis of The Gaussian Wiretap Multiple-Access Channel}
\author{\IEEEauthorblockN{Parisa Babaheidarian*, Somayeh Salimi**, Panos Papadimitratos**}
\IEEEauthorblockA{
*Boston University,**KTH Royal Institute of Technology}}

\date{}
\maketitle
\begin{abstract}
In this work, we consider a $K$-user Gaussian wiretap multiple-access channel (GW-MAC) in which each transmitter has an independent confidential message for the receiver. There is also an external eavesdropper who intercepts the communications. The goal is to transmit the messages reliably while keeping them confidential from the eavesdropper. To accomplish this goal, two different approaches have been proposed in prior works, namely, i.i.d. Gaussian random coding and real alignment. However, the former approach fails at moderate and high SNR regimes as its achievable result does not grow with SNR. On the other hand, while the latter approach gives a promising result at the infinite SNR regime, its extension to the finite-SNR regime is a challenging task. To fill the gap between the performance of the existing approaches, in this work, we establish a new scheme in which, at the receiver's side, it utilizes an extension of the compute-and-forward decoding strategy and at the transmitters' side it exploits lattice alignment, cooperative jamming, and i.i.d. random codes. For the proposed scheme, we derive a new achievable bound on sum secure rate which scales with $\log(\mathrm{SNR})$ and hence it outperforms the i.i.d. Gaussian codes in moderate and high SNR regimes. We evaluate the performance of our scheme, both theoretically and numerically. Furthermore, we show that our sum secure rate achieves the optimal sum secure degrees of freedom in the infinite-SNR regime.
\end{abstract}
\section{Introduction}\label{sec1}
It has been shown that structured codes outperform the standard random codes in certain communication scenarios with and without security constraints such as~\cite{korner1979encode}, \cite{bagherikaram2010secure},\cite{xie2014secure}, and \cite{he2014providing}. For instance, a scheme based on real alignment was proposed in~\cite{bagherikaram2010secure} for the $K$-user Gaussian wiretap multiple-access channel (GW-MAC). They used alignment to confuse the eavesdropper and showed that in the infinite SNR regime, their scheme improves over the result achieved by Gaussian i.i.d. random codes in~\cite{tekin2008general}. Also, Xie \textit{et al.} in~\cite{xie2014secure} employed the real alignment technique in conjunction with cooperative jamming for the same channel model and showed improvement over the former scheme in~\cite{bagherikaram2010secure}. In particular, the proposed scheme in~\cite{xie2014secure} achieves the optimal sum secure degrees of freedom (s.d.o.f.) for the $K$-user Gaussian wiretap multiple-access channel.  Additionally, using similar schemes, they characterized the exact secure degrees of freedom for Gaussian broadcast channel with multiple helpers as well as sum secure degrees of freedom for the two-user interference channel.\\
\indent The aforementioned alignment schemes are limited to the infinite SNR regime whereas their extension to the finite-SNR regime is challenging. On the other hand, in~\cite{ordentlich2014approximate}, Ordentlich  \textit{et al.} accommodated the finite-SNR regime analysis for the $K$-user multiple-access channel (MAC) without security constraints. They proposed a lattice alignment scheme using the compute-and-forward decoding strategy, introduced in~\cite{nazer2011compute}, and showed that their scheme achieves a sum rate, \textit{without security}, that is within a constant gap from $K$-user MAC sum capacity and is valid for any finite value of SNR. \\
\indent Also, very recently, a compute-and-forward based scheme was proposed to handle the finite-SNR regime for the $K$-user Gaussian wiretap multiple-access channel~\cite{paper1}. A lower bound on the sum secure rate was derived, which achieved $\frac{K-1}{K}$ sum secure degrees of freedom. This was the first scheme on GW-MAC that achieved a positive s.d.o.f and yet it worked at finite-SNR regime. \\
\indent In this paper, in light of the work in~\cite{xie2014secure}, we further improve upon~\cite{paper1} such that our new achievable sum secure rate reaches the optimal sum secure degrees of freedom, i.e., $\frac{K(K-1)}{K(K-1)+1}$, in the infinite-SNR regime and yet improves over i.i.d. Gaussian random codes in the moderate and high SNR regimes. It also surpasses the result in~\cite{paper1} in high SNR regimes. We propose a new scheme which consists of two layers in its encoding strategy: the inner layer and the outer layer. Transmitters incorporate a nested lattice structure as well as cooperative jamming signals in their inner layers and i.i.d. random codes in their outer layers. Also, in our decoding strategy, the receiver exploits a new extension of the compute-and-forward decoding strategy. We characterize a lower bound on the sum secure rate for the $K$-user Gaussian wiretap multiple access channel which is valid for any finite value of SNR and is in agreement with the result in~\cite{xie2014secure} in the infinite-SNR regime. Moreover, we evaluate the performance of our proposed scheme numerically for a three-user GW-MAC.\\
\indent The rest of the paper is organized as follows. In Section~\ref{sec2} the system model is defined. Section~\ref{sec3} is devoted to our main results. The achievability scheme and analysis of security are presented in Section~\ref{sec4}. Section~\ref{sec5} provides the numerical results. The paper is concluded in Section~\ref{sec6}. Finally, complementary proofs are provided in Appendix.
\section{System model} \label{sec2}
We consider the problem of secure and reliable communication over a multiple-access channel with $K$ users at the presence of an external eavesdropper. The system is modeled by
\begin{equation} \label{eq1}
\mathbf{y}=\sm h_{\ell}\mathbf{x}_{\ell}+\mathbf{z}, \quad \mathbf{y}_E=\sm g_{\ell}\mathbf{x}_{\ell}+\mathbf{z}_E
\end{equation}
Where $\mathbf{x}_{\ell}$ is user $\ell$'s channel input with a block length of $N$. Vectors $\mathbf{y}$ and $\mathbf{y}_E$ are the channel outputs at the receiver and the eavesdropper sides, respectively. The real-valued elements $h_{\ell}$ and $g_{\ell}$ are the channel gains from user $\ell$ to the receiver and the eavesdropper, respectively; thus, vectors $\mathbf{h}\triangleq[h_1,\dots,h_K]^{T}$ and $\mathbf{g}\triangleq[g_1,\dots,g_K]^{T}$ are the channel gain vectors. We assume that the transmitters\footnote{In our scheme, knowledge of the channel state is not beneficiary either to the receiver or the eavesdropper.} know the channel states, i.e., the channel gain vectors, in advance. Finally, the random vectors $\mathbf{z}$ and $\mathbf{z}_E$ are, respectively, the receiver's and the eavesdropper's channel noises which are independent and each is i.i.d. Gaussian with zero mean and normalized variance.\\
\indent As it is shown in Figure 1, user~$\ell$ has an independent confidential message $W_{\ell}$ which is uniformly distributed over the set $\{1,\dots, 2^{NR_{\ell}}\}$, for $\ell \in~\{1, \dots, K\}$. User $\ell$ maps its message to the codeword $\mathbf{x}_{\ell}$ through a stochastic encoder, i.e., $\mathbf{x}_{\ell} = \mathcal{E}_{\ell}(W_{\ell})$. Also, there are power constraints on the channel inputs as $\|\mathbf{x_{\ell}}\|^2 \leq NP$, for all $\ell \in\{1,\dots,K\}$. There is also a decoder $\mathcal{D}$ at the receiver side that estimates the transmitted messages, i.e., $\mathcal{D}(\mathbf{y})=\{\hat{W}_{\ell}\}_{\ell=1}^{K}$.
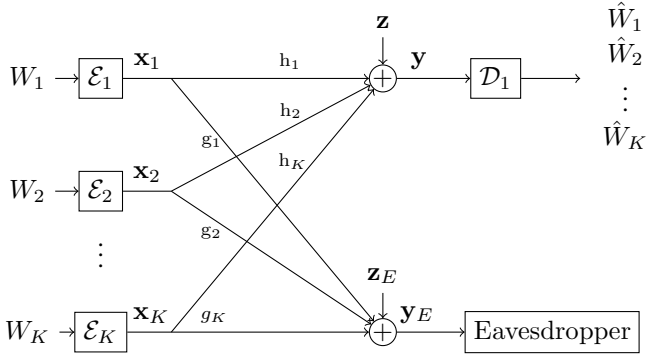
\begin{figure}
\centering
\begin{tikzpicture}[scale=0.75]
\draw(0.7,6) node (nodew1) {$W_1$};
\draw (2,6) node[draw] (nodeE1) {$\mathcal E_1$};
\draw (7,6) node[circle,draw,inner sep=0] (nodeplus1) {$+$};
\draw (7,7) node (nodenoise1) {$\mathbf z$};
\draw (9,6) node[draw] (nodeD1) {$\mathcal D_1$};

\draw(11.3,6) node (nodewhat1) {$\begin{array}{c}
 \hat W_1 \\ \hat W_2 \\ \vdots \\ \hat W_K \end{array} $
 };
\draw[->] (nodew1)--(nodeE1);
\draw[->] (nodeE1)--(nodeplus1) node[pos=0.1,sloped,above]{$\mathbf{x}_1$} node[pos=0.68,sloped,above] {$\scriptstyle \mathrm h_1$};
\draw[->] (nodenoise1)--(nodeplus1);
\draw[->] (nodeplus1)--(nodeD1) node[pos=0.3,sloped,above]{$\mathbf y$};
\draw[->] (nodeD1)--(nodewhat1);
\draw(0.7,4) node (nodew2) {$W_2$};
\draw (2,4) node[draw] (nodeE2) {$\mathcal E_2$};
\draw[->] (nodew2)--(nodeE2);
\draw (nodeE2)--(3.25,4) node[pos=0.5,sloped, above]{$\mathbf{x}_2$};
\draw(0.7,1.5) node (nodew3) {$W_K$};
\draw (2,1.5) node[draw, minimum size = 10] (nodeE3) {$\mathcal E_K$};
\draw (7,1.5) node[circle,draw,inner sep=0] (nodeplus3) {$+$};
\draw (7,2.5) node (nodenoise3) {$ \mathbf{z}_E$};
\draw (10,1.5) node[draw] (nodeD3) {$ \mathrm {Eavesdropper}$};
\draw[->] (nodew3)--(nodeE3);
\draw[->] (nodeE3)--(nodeplus3)node[pos=0.1,sloped,above]{$\mathbf{x}_K$} node[pos=0.36,sloped,above] {$\scriptstyle g_K$};
\draw[->] (nodenoise3)--(nodeplus3);
\draw[->] (nodeplus3)--(nodeD3) node[pos=0.3,sloped,above]{$\mathbf y_E$};
\draw (2,3) node {$\vdots$};
\draw[->] (3.25,6)--(nodeplus3) node[pos=0.2,below]{$\scriptstyle \mathrm g_1$};

\draw[->] (3.25,4)--(nodeplus1) node[pos=0.6,above]{$\scriptstyle \mathrm h_2$};
\draw[->] (3.25,4)--(nodeplus3) node[pos=0.2,below]{$\scriptstyle \mathrm g_2$};
\draw[->] (3.25,1.5)--(nodeplus1) node[pos=0.6,above, yshift=3]{$\scriptstyle \mathrm h_K$};
\end{tikzpicture}
\vspace{0.02 in}
\caption{ \small {The Gaussian wiretap multiple-access channel model.}}
\vspace{-5mm}
\end{figure}
\begin{defi}[Achievable sum secure rate]\label{achievability tuple}
For~the $K$-user GW-MAC, a sum secure rate~$\sm R_{\ell}$ is achievable, if for any$~\epsilon>0$ and sufficiently large $N$, there exist a sequence of encoders $\{\mathcal{E}_{\ell}\}_{\ell=1}^{K}$ and a decoder $D$ such that:
\vspace{-2.6mm}
\begin{equation}
\mathrm{Prob}\left(\bigcup_{\ell=1}^K\lbrace \hat{W}_{\ell} \neq W_{\ell} \rbrace\right)< \epsilon \label{eq2}
\end{equation}
\vspace{-2.5mm}
\begin{equation}
\sm R_{\ell}\leq \frac{1}{N}H(W_1,W_2,\dots,W_K|\mathbf{y}_E)+\epsilon \label{eq3}
\end{equation}
\end{defi}
We refer to inequalities (\ref{eq2}) and (\ref{eq3}) as the \textit{reliability} and the \textit{weak secrecy} constraints, respectively. Note that the sum secure capacity is the supremum over all the achievable sum secure rates.
\section{Main results}\label{sec3}
In our achievability scheme, we develop a new decoding strategy for $K$-user GW-MAC, which extends the one used in~\cite{ordentlich2014approximate}, to comply with our encoding strategy. In our setting, the receiver decodes $K(K-1)+1$ equations whose coefficient vectors are integer valued and are linearly independent, then it solves the system of the equations for the transmitted messages. We denote the optimal rates at which the receiver can successfully decode the equations by the set of rates $R_{comb,k},~\forall~k\in \{1,\dots,K(K-1)+1\}$ and we refer to them as equation rates or integer combination rates. The rates $R_{comb,k}$ are computed in Section~\ref{sec4}.\\
\indent The following theorem is our main result.
\begin{theorem}
A sum secure rate $\sm R_{\ell}$ is achievable if it satisfies the following inequality
\begin{IEEEeqnarray}{l}
\nonumber \sm R_{\ell}<
\sum_{k=2}^{K(K-1)+1} R_{comb,k}
-\\ \frac{1}{2} \sum_{\ell=1}^K \log \left(\frac{\sum_{j=1, j \neq \ell}^{K} \gamma^2_{j,\ell}P_{j,\ell}+\gamma^2_{\ell,\ell}P_{\ell,\ell}}{\gamma^2_{\ell,\ell}P_{\ell,\ell}}\right) \label{th2}
\end{IEEEeqnarray}\\
such that, for all $\ell \in \{1,\dots,K\}$, the following holds.
\begin{equation}\label{align1new}
\gamma_{(\ell,i)}= \frac{g_{\ell}h_i}{g_i}~~\forall i\neq \ell,
\end{equation}
\begin{equation} \label{align2new}
\gamma_{(\ell,\ell)}= h_{\ell}
\end{equation}
and
\begin{equation}\label{poweraloc}
\sum_{i=1}^K P_{\ell,i}\leq P,~P_{\ell,i}>0~\forall i\in\{1,\dots,K\}
\end{equation}
\end{theorem}
Note that the achievable bound in~(\ref{th2}) is a function of coefficients $P_{\ell,i}$, hence the supremum over all choices of $P_{\ell,i}$ satisfying~(\ref{poweraloc}) is also achievable.\\
\begin{remark}
The sum rate $\medmuskip=0mu
\thinmuskip=0mu
\thickmuskip=0mu \sum_{k=1}^{K(K-1)+1}R_{comb,k}$ in our scheme is different from the sum rate achieved by~\cite{ordentlich2014approximate} for the $K$-user Gaussian MAC capacity\footnote{We refer to the sum of non-secure rate as sum rate.}. However, we show that our achievable sum rate reaches the performance of the compute-and-forward framework in~\cite{ordentlich2014approximate}, asymptotically. In other words, we show that $\sum_{k=1}^{K(K-1)+1}R_{comb,k}$ is within a constant gap from $K$-user Gaussian MAC capacity. The proof is provided in Appendix-A. Also, we will verify this claim by a numerical experiment in Section~\ref{sec5}.
\end{remark}
\begin{corollary}
The achievable sum secure rate in (\ref{th2}) grows with $\log(\mathrm{SNR})$, i.e.,
\begin{equation} \label{cornew}
\sm R_{\ell} \propto \log(\mathrm{P})
\end{equation}
\end{corollary}
\begin{remark}
The importance of Corollary~1 becomes clear when its result is compared with the performance of sum secure rate provided by i.i.d. Gaussian random codes in~\cite{tekin2008general}. Recall that the latter achieves a sum secure rate of $\frac{1}{2}\log(\frac{1+\|\mathbf{h}\|^2P}{1+\|\mathbf{g}\|^2P})$ for the $K$-user GW-MAC. Hence, the security performance of the scheme in~\cite{tekin2008general} does not grow by increasing the power\footnote{Assuming the channel gain vector norms are of similar orders.}.
\end{remark}
\begin{corollary}
The sum secure degrees of freedom (s.d.o.f.) achieved by our scheme is 
\begin{equation}
\lim_{P\rightarrow \infty} \frac{\sum_{\ell} R_{\ell}}{\frac{1}{2}\log \left(1+P\right)}=\frac{K(K-1)}{K(K-1)+1}
\end{equation}
\end{corollary}
\begin{remark} Recall that in~\cite{xie2014secure} it is shown that the optimal sum secure degrees of freedom for the $K$-user Gaussian wiretap MAC is $\frac{K(K-1)}{K(K-1)+1}$. Hence, our achievable result in~(\ref{th2}) is asymptotically optimal.
\end{remark}

\indent Next section provides the achievability scheme and the security analysis.
\section{The achievability scheme and analysis of security}\label{sec4}
In this section, we introduce our new scheme. First, we describe the codebook construction and encoding strategy applied by the transmitters and then we unfold the decoding strategy used by the receiver. Finally, the proof of weak secrecy of the proposed scheme is presented. (Note: Proofs for Corollary~1 and 2 are given in Appendix.
\subsection*{Codebook construction and encoding strategy}
In order to send the confidential messages $\{W_{\ell}\}_{\ell=1}^K$, each user (transmitter) utilizes $(K-1)$ sub-codewords to encode its own confidential message. Moreover, each user employs an additional sub-codeword as a jamming signal which does not carry any information regarding the confidential messages. We proceed with describing the operations done by user $\ell$, other users perform similarly. \\
\indent User~$\ell$ picks $K$ $n$-dimensional coarse and fine lattice pairs as $(\Lambda_{f,(\ell,i)},\Lambda_{(\ell,i)})$ for all $i \in \{1,\dots,K\}$. The set of lattices used by all users form a nested structure in which the following two conditions hold.
\begin{equation}\label{condition1}
\Lambda_{(j,\ell)} \subseteq \Lambda_{(\ell,\ell)},~~
\Lambda_{f,(j,\ell)} \subseteq \Lambda_{f,(\ell,\ell)}~~\forall~j\neq \ell
\end{equation}
The fundamental Voronoi region of the coarse lattice~$\Lambda_{(\ell,i)}$ is denoted by $\mathcal{V}_{(\ell,i)}$. For each $i$, the centers of the translations of the fine lattice $\Lambda_{f,(\ell,i)}$ (cosets) lying in $\mathcal{V}_{(\ell,i)}$ are considered as the realizations of the random vector $\mathbf{t}_{\ell,i}$. The second moment of the coarse lattice $\Lambda_{(\ell,i)}$ is set as $\gamma^2_{(\ell,i)}P_{\ell,i}$.\\
\indent Define the set $\mathcal{L}_{(\ell,i)}\triangleq\{\mathbf{t}_{\ell,i}|\mathbf{t}_{\ell,i}\in  \mathcal{V}_{(\ell,i)}\}$. Assume that vectors $\mathbf{t}_{\ell,i}$ have a probability distribution $P(\mathbf{t}_{\ell,i})$ over the set $\mathcal{L}_{(\ell,i)}$. The set $\mathcal{L}_{(\ell,i)}$ is termed the \textit{inner} sub-codebook~$i$ used by user~$\ell$. \\
\indent Consider a one-to-one mapping $\phi:\{1,\dots,K\}\times \{1,\dots,K\}\backslash \{(\ell,\ell)|\ell \in [1,K]\} \rightarrow \{2,\dots,K(K-1)+1\}$. Then, the ratio between the coarse lattice $\Lambda_{(\ell,i)}$ and its associated fine lattice is set such that $R_{comb,k}=\frac{1}{n}\log(|\mathcal{L}_{(\ell,i)}|)$ where $k=\phi(\ell,i)$ for all $i \neq \ell$ and $k=1$ for $i=\ell$. \\
\indent Next, user $\ell$ generates $B$ independent copies of vectors $\mathbf{t}_{\ell,i}$ according to distribution $P(\mathbf{t}_{\ell,i})$ to make one \textit{realization} of the outer codewords $ \bar{\mathbf{t}}_{\ell,i}$. As a result, the block length of the generated outer codeword is $N\triangleq n \times B$. User $\ell$ performs this procedure $2^{NR_{comb,(\ell,i)}}$ times to construct its outer sub-codebook~$i$, i.e., $\mathcal{C}_{(\ell,i)}$. User $\ell$ generates all its other sub-codebooks, similarly. It is worth to mention that the i.i.d. repetition of the inner codebook is added to the scheme so that we can benefit from Packing Lemma in the proof of weak secrecy\footnote{Recall that proof of Packing Lemma is followed from jointly typicality lemma on i.i.d. random sequences.}.\\
\indent In the next step, similar to the Wyner random partitioning in~\cite{wyner}, the outer codewords of each sub-codebook~$(\ell,i)$, for all $\ell \in \{1,\dots,K\}$ and $i\neq \ell$, are randomly partitioned into $2^{NR_{\ell,i}}$ bins of equal sizes. Note that the non-negative rates $R_{\ell,i}$ are chosen by user $\ell$ such that $\sum_{i\neq \ell} R_{\ell,i}= R_{\ell}$. The partition index is characterized by a random variable $W_{\ell,i}$ in the corresponding sub-codebook. To encode the confidential message $W_{\ell}$ with its realizations $w_{\ell}\in \{1,\dots,2^{NR_{\ell}}\}$, user $\ell$ divides its message into $(K-1)$ mutually independent sub-messages $W_{\ell,i}$ with the corresponding realizations $w_{\ell,i}$. For the sub-message $i$, $i \in \{1,\dots,K\}\backslash {\ell}$, the user picks randomly a codeword $ \bar{\mathbf{t}}_{\ell,i}$ from the partition $w_{\ell,i}$ in the sub-codebook $\mathcal{C}_{(\ell,i)}$\footnote{Once the assignment of the sub-messages to codewords is done, it will be fixed and provided to all parties.}. Then, it adds a random dither vector $ \bar{\mathbf{d}}_{\ell,i}$ to the selected codeword\footnote{Each $n$-length block of the dither $ \bar{\mathbf{d}}_{(\ell,i)}$ is uniformly distributed over $\mathcal{V}_{(\ell,i)}$.} and reduces the sum modulo the coarse lattice $\Lambda_{(\ell,i)}$. Keep in mind that dithers are independent of all other variables and are public. \\
\indent The modular operation is done for each block of size $n$ and the outcomes of all $B$ blocks are concatenated together. We have:
\begin{equation}
\tilde{\mathbf{x}}_{\ell,i}\triangleq\left(\left[\bar{\mathbf{t}}_{\ell,i}+\bar{\mathbf{d}}_{\ell,i}\right] \modnli\right)
\end{equation}
 At the end, the resulting codeword is scaled by $\frac{1}{\gamma_{(\ell,i)}}$. In short, the $N$-length codeword $\mathbf{x}_{\ell,i}$ is constructed as
\begin{equation}
\mathbf{x}_{\ell,i}\triangleq \frac{1}{\gamma_{(\ell,i)}}\left(\left[\bar{\mathbf{t}}_{\ell,i}+\bar{\mathbf{d}}_{\ell,i}\right] \modnli\right)
\end{equation}
The scaling factors $\gamma_{(\ell,i)}$ are defined in (\ref{align1new}) and (\ref{align2new}).\\
\indent In addition, the jamming codeword $\bar{\mathbf{t}}_{\ell,\ell}$ is chosen uniformly at random from the sub-codebook $\mathcal{C}_{(\ell,\ell)}$. The codeword $\mathbf{x}_{\ell,\ell}$ is constructed similarly. Eventually, the superposition codeword $\mathbf{x}_{\ell}\triangleq \sum_{i=1}^K \mathbf{x}_{\ell,i}$ is transmitted through the channel by user~$\ell$. This procedure is displayed in Figure 2(a).
\begin{figure}
\centering
\begin{tikzpicture}[scale=0.75]
\draw(0.5,6) node (nodew1) {$W_{\ell}$};
\draw (2.6,5.95) node (nodewli) {$\{W_{\ell,i}\}_{i=1 \atop i\neq \ell}^K$};
\draw[->] (nodew1)--(nodewli);
\draw (4.5,6) node[draw] (nodeE1) {$\mathcal E_{\ell}$};
\draw[->] (nodewli)--(nodeE1);
\draw (4.5,7) node (nodejam){$\mathrm{Jamming~signal}$};
\draw[->] (nodejam)--(nodeE1);
\draw (6.4,6) node (nodetli) {$\{\bar{\mathbf{t}}_{\ell,i}\}_{i=1}^K$};
\draw[->] (nodeE1)--(nodetli);
\draw (7.5,6) node[circle,draw,inner sep=0] (nodeplus1) {$+$};
\draw (7.5,7) node (noded){$\{\bar{\mathbf{d}}_{\ell,i}\}_{i=1}^K$};
\draw[->] (noded)--(nodeplus1);
\draw (9,6) node[draw] (nodemod){$\mathrm{mod}$};
\draw[->] (nodeplus1)--(nodemod);
\draw (11,6) node (nodextildli) {$\{\tilde{\mathbf{x}}_{\ell,i}\}_{i=1}^K$};
\draw[->] (nodemod)--(nodextildli);
\draw (2,5) node (nodextildli2) {$\{\tilde{\mathbf{x}}_{\ell,i}\}_{i=1}^K$};
\draw (3.6,5) node[circle,draw,inner sep=0] (nodetimes1) {$\times$};
\draw[->] (nodextildli2)--(nodetimes1);
\draw (3.6,4) node (nodefactor){$\frac{1}{\gamma_{\ell,i}}$};
\draw[->] (nodefactor)--(nodetimes1);
\draw (5.5,5) node (nodexli) {$\{\mathbf{x}_{\ell,i}\}_{i=1}^K$};
\draw[->] (nodetimes1)--(nodexli);
\draw (7.5,5) node[draw] (nodesum) {$\sum_i$};
\draw[->] (nodexli)--(nodesum);
\draw (9,5) node (nodexl) {$\mathbf{x}_{\ell}$};
\draw[->] (nodesum)--(nodexl);
\draw (5.5,3.3) node (nodefig2a) {\small{Figure 2(a): Encoder~$\ell$.}};
\end{tikzpicture}
\begin{tikzpicture}[scale=0.75]
\draw(0.5,2) node (nodey) {$\mathbf{y}$};
\draw (1.5,2) node[circle,draw,inner sep=0] (nodetimes2){$\times$};
\draw[->] (nodey)--(nodetimes2);
\draw (1.5,1) node (nodebeta){$\beta_{\mathrm{MSE}}$};
\draw[->] (nodebeta)--(nodetimes2);
\draw (3.5,2) node[draw] (noded2) {$\mathrm{Dither} \atop \mathrm{subtraction}$};
\draw[->] (nodetimes2)--(noded2);
\draw(6,2) node[draw] (nodemod2) {$\mathrm{mod}$};
\draw[->] (noded2)--(nodemod2);
\draw (8.5,2) node[draw] (nodeQ) {$\mathrm{Lattice} \atop \mathrm{qunatization}$};
\draw[->] (nodemod2)--(nodeQ);
\draw (11,2) node (nodevk) {$\hat{\mathbf{v}}_{k}$};
\draw[->] (nodeQ)--(nodevk);
\draw (5.5,0.5) node (nodefig2b) {\small{Figure 2(b): The Decoder.}};
\end{tikzpicture}
\vspace{0.02 in}
\caption{ \small {Our achievable scheme.}}
\vspace{-5mm}
\end{figure}
\subsection*{Decoding strategy}
\indent Recall that the scaling factors $\gamma_{\ell,\ell},~\forall\ell$, are chosen such that the users' jamming codewords $\tilde{\mathbf{x}}_{\ell,\ell}$ get aligned at the receiver's side and form a single lattice codeword $\sum_{\ell}\tilde{\mathbf{x}}_{\ell,\ell}$. Thus, the receiver observes
\begin{IEEEeqnarray}{L}\label{align3}
\nonumber \mathbf{y}=\sum_{\ell}\sum_{i \atop i \neq \ell} \frac{h_{\ell}g_i}{g_{\ell}h_i} \left(\left[\bar{\mathbf{t}}_{\ell,i}+\bar{\mathbf{d}}_{\ell,i}\right] \modnli\right)\\
+\sum_{\ell} \left[\bar{\mathbf{t}}_{\ell,\ell}+\bar{\mathbf{d}}_{\ell,\ell}\right] \modnll,
\end{IEEEeqnarray}
\indent Extending the compute-and-forward decoding strategy in~\cite{nazer2011compute} to $K(K-1)+1$ integer combinations, the receiver estimates the sub-messages codewords $\bar{\mathbf{t}}_{\ell,i}$ by decoding $K(K-1)+1$ linearly independent equations whose coefficients are integer valued. Equation $k$ is denoted by $\mathbf{v}_k$ and it's defined as
\begin{equation}
\medmuskip=0mu
\thinmuskip=0mu
\thickmuskip=0mu
\mathbf{v}_{k}\triangleq\left[a^{(k)}_{1}\sum_{\ell}\bar{\mathbf{t}}_{\ell,\ell}+\sum_{(\ell,i) \atop (\ell \neq i)}a^{(k)}_{\ell,i}\bar{\mathbf{t}}_{\ell,i}\right]\modn,
\end{equation}
for all $k\in\{1,\dots,K(K-1)+1\}$. Thus, the integer coefficients vector for equation $k$ is a $(K(K-1)+1)\times 1 $ vector $\mathbf{a}^{(k)}$. \\
\indent To decode equation $\mathbf{v}_{k}$, the receiver scales its observation $\mathbf{y}$ by a factor of $\beta$ and then it subtracts the dithers off, finally it reduces the result modulo lattice $\Lambda$. It is worth mentioning that $\Lambda$ is the coarsest lattice among all the previously defined lattice sets. The decoding procedure is depicted in Figure~2(b). We have
\begin{IEEEeqnarray}{l}
\medmuskip=0mu
\thinmuskip=0mu
\thickmuskip=0mu
\nonumber \mathbf{s}_k=\left[\beta \mathbf{y}-\sum_{(\ell,i) \atop i \neq \ell }a^{(k)}_{\ell,i}\bar{\mathbf{d}}_{\ell,i}-a^{(k)}_{1}\sum_{\ell}\bar{\mathbf{d}}_{\ell,\ell}\right]\modn\\ 
\medmuskip=0mu
\thinmuskip=0mu
\thickmuskip=0mu
\nonumber 
\begin{split}
=\left[\sum_{(\ell,i) \atop i \neq \ell}a^{(k)}_{\ell,i}\gamma_{\ell,i}\mathbf{x}_{\ell,i}+a^{(k)}_{1}\sum_{\ell}\gamma_{\ell,\ell}\mathbf{x}_{\ell,\ell}-\sum_{(\ell,i)\atop i \neq \ell }a^{(k)}_{\ell,i}\bar{\mathbf{d}}_{\ell,i} \right. \\
\left. -a^{(k)}_{1}\sum_{\ell}\bar{\mathbf{d}}_{\ell,\ell}+
\mathbf{z}_{\mathrm{eff},k}(\mathbf{h},\vec{\gamma},\mathbf{a}^{(k)},\beta)\right]\modn
\end{split} 
 \\
=\left[\mathbf{v}_{k}+\mathbf{z}_{\mathrm{eff},k}(\mathbf{h},\vec{\gamma},\mathbf{a}^{(k)},\beta) \right] \modn
\end{IEEEeqnarray}
Where $\vec{\gamma}$ is the vector of the scaling factors and $\mathbf{z}_{\mathrm{eff},k}$ represents the effective noise defined as
\begin{IEEEeqnarray}{l}\label{noiseef}
\medmuskip=0mu
\thinmuskip=0mu
\thickmuskip=0mu
\nonumber 
\mathbf{z}_{\mathrm{eff},k}(\mathbf{h},\vec{\gamma},\mathbf{a}^{(k)},\beta) \triangleq(\beta-a^{(k)}_{1})\sum_{\ell}\gamma_{\ell,\ell}\mathbf{x}_{\ell,\ell}\\+ \sum_{(\ell,i) \atop i \neq \ell}(\beta h_{\ell}-a^{(k)}_{\ell,i}\gamma_{\ell,i})\mathbf{x}_{\ell,i}
+\beta \mathbf{z}
\end{IEEEeqnarray}
Thus, the effective noise variance is given as
\begin{equation}\label{effdef}
\sigma^2_{\mathrm{eff},k}=\|\beta\tilde{\mathbf{h}}-\Gamma.\mathbf{a}^{(k)}\|^2P+\beta^2
\end{equation}
$\tilde{\mathbf{h}}$ is a $(K(K-1)+1) \times 1$ vector defined as
\begin{equation}
\tilde{\mathbf{h}}\triangleq [1,\underbrace{h_1,\dots,h_1}_{K-1},\underbrace{h_2,\dots,h_2}_{K-1},\dots \dots,\underbrace{h_K,\dots,h_K}_{K-1}]^T
\end{equation}
$\Gamma$ is a $(K(K-1)+1) \times (K(K-1)+1)$ diagonal matrix such that:
\begin{equation}
\Gamma(1,1)=\sqrt{\frac{\sum_{\ell=1}^K\gamma_{\ell,\ell}^2P_{\ell,\ell}}{P}}
\end{equation}
and for $k>2$
\begin{equation}
\Gamma(k,k)=\sqrt{\frac{\gamma_{\ell,i}^2P_{\ell,i}}{P}}~~s.t.~k=\phi(\ell,i)
\end{equation}
It can be shown that the minimizer of the effective noise variance with respect to $\beta$ is the MSE factor in linear estimation of $a^{(k)}_1\sum_{\ell}\gamma_{\ell,\ell}\mathbf{x}_{\ell,\ell}+\sum_{(\ell,i) \atop i \neq \ell}a^{(k)}_{\ell,i}\gamma_{\ell,i}\mathbf{x}_{\ell,i}$ from vector $\mathbf{y}$. Therefore, substituting the optimal $\beta$ in equation~(\ref{effdef}) yields
\begin{equation}
\sigma^2_{\mathrm{eff},k}(\mathbf{h},\vec{\gamma},\mathbf{a}^{(k)})=\|\mathbf{F}.\mathbf{a}^{(k)}\|^2
\end{equation}
Where the dot operation is the matrix-vector product and matrix $\mathbf{F}$ is given as
\begin{equation}
\mathbf{F}\triangleq\big(\frac{1}{P}\Gamma^{-2}+\tilde{\mathbf{h}}\tilde{\mathbf{h}}^T\big)^{\frac{-1}{2}}.\Gamma
\end{equation}
\indent Note that among the lattice codewords participating in equation $\mathbf{v}_k$, the one constructed on the densest lattice can be recovered by decoding the equation $\mathbf{v}_k$. The receiver decodes upon receiving each $n$-length block and concatenates the block estimates at the end to get an estimate of the transmitted outer codeword. Consider a $k>2$ and assume that lattice $\Lambda_{f,(\ell,i)}$ is the densest lattice participating in $\mathbf{v}_k$. In fact, mapping $\phi$ is deduced from this step, i.e., $k=\phi(\ell,i)$. Subsequently, the receiver estimates the corresponding codeword $\bar{\mathbf{t}}_{\ell,i}$ by decoding the equation $\mathbf{v}_k$ as
\begin{equation}\label{decodingvk}
\hat{\mathbf{v}}_k=\left[Q_{f,(\ell,i)}(\mathbf{s}_k)\right] \modn
\end{equation}
Where $Q_{\Lambda}(.)$ is the nearest neighbor qunatizer associated with lattice $\Lambda$. From (\ref{decodingvk}), it can be seen that the probability of decoding equation $\mathbf{v}_k$ with error is upper-bounded as
\begin{equation*}
\mathrm{Prob}(\hat{\mathbf{v}}_k \neq \mathbf{v}_k)=\mathrm{Prob}(\mathbf{z}_{\mathrm{eff},k}\notin \mathcal{V}_{(\ell,i)})
\end{equation*} 
\indent Consequently, based on arguments similar to those in Theorem~2 in~\cite{ordentlich2014approximate}, it can be shown that for sufficiently high dimensional lattices, the decoding error probability can be chosen smaller than $\epsilon$, for any arbitrary $\epsilon>0$, provided that
\begin{equation}
R_{comb,k}<\frac{1}{2}\log\left(\frac{\gamma_{(\ell,i)}^2P_{\ell,i}}{\sigma^2_{\mathrm{eff},k}}\right)~~s.t.~\phi(\ell,i)=k,~k \neq 1
\end{equation}
and
\begin{equation}
R_{comb,1}<\max_{\ell}\left(\frac{1}{2}\log\left(\frac{\gamma_{(\ell,\ell)}^2P_{\ell,\ell}}{\sigma^2_{\mathrm{eff},1}}\right)\right)
\end{equation}
Finally, recall that by Theorem~1, the achievable sum secure rate must satisfy in $\sm R_{\ell}<\sum_{k=2}^{K(K-1)+1}R_{comb,k}$; therefore we have
\begin{equation*}
\mathrm{Prob}\left(\bigcup_{\ell=1}^K\lbrace \hat{W}_{\ell} \neq W_{\ell} \rbrace\right)< \epsilon
\end{equation*}
\subsection*{Proof of weak secrecy}
Base on our scheme, the eavesdropper observes the sequence $\mathbf{y}_E$ as
 \begin{IEEEeqnarray}{l}
\nonumber  \mathbf{y}_E=\sum_{\ell =1}^K \sum_{i=1 \atop i \ne \ell}^K \frac{g_i}{h_i}\left(\left[\bar{\mathbf{t}}_{\ell,i}+\bar{\mathbf{d}}_{\ell,i}\right]\modnli\right)\\
+\sum_{\ell =1}^K\left( \frac{g_{\ell}}{h_{\ell}}\left[\bar{\mathbf{t}}_{\ell,\ell}+\bar{\mathbf{d}}_{\ell,\ell}\right]\modnll\right)+\mathbf{z}_E
 \end{IEEEeqnarray}
 \vspace{-2mm}
\indent Define a sequence of vectors $\mathbf{y}_E^{(1)},\dots,\mathbf{y}_E^{(K)}$ where
\begin{IEEEeqnarray}{l}
\nonumber  \mathbf{y}_E^{(\ell)}\triangleq\ \sum_{j=1 \atop j \ne \ell}^K \left(\left[\bar{\mathbf{t}}_{j\ell}+\bar{\mathbf{d}}_{j,\ell}\right]\modnjl\right)\\
+\left( \left[\bar{\mathbf{t}}_{\ell,\ell}+\bar{\mathbf{d}}_{\ell,\ell}\right]\modnll\right)
 \end{IEEEeqnarray}
 Note that
\begin{IEEEeqnarray*}{l}
\medmuskip=0mu
\thinmuskip=0mu
\thickmuskip=0mu \frac{1}{N}H(W_1,\dots, W_K \big |\mathbf{y}_E)\geq
 \frac{1}{N}H(W_1,\dots, W_K\big | \mathbf{y}_E^{(1)},\dots,\mathbf{y}_E^{(K)}).
 \end{IEEEeqnarray*}
\indent First, we show that the following sum secure rate provides weak secrecy.
\begin{equation} \label{sumrate1}
\medmuskip=0mu
\thinmuskip=0mu
\thickmuskip=0mu
\sm R_{\ell}=\frac{1}{N}H\left(\bar{\mathbf{t}}_{1,1},\bar{\mathbf{t}}_{1,2}, \dots , \bar{\mathbf{t}}_{K-1,K},\bar{\mathbf{t}}_{K,K}\big|\mathbf{y}_E^{(1)},\dots,\mathbf{y}_E^{(K)},D\right)+\epsilon_1
\end{equation} 
In which $\epsilon_1>0$ goes to zero as $N \rightarrow \infty$ and $D$ is defined as $D\triangleq\{\bar{\mathbf{d}}_{\ell,i}\}_{(\ell,i)}$. We have
\begin{IEEEeqnarray}{L} 
\nonumber \frac{1}{N}I(W_1,\dots , W_K;\mathbf{y}_E|D) \leq \\\nonumber
\medmuskip=0mu
\thinmuskip=0mu
\thickmuskip=0mu
 \frac{1}{N}H(W_1,\dots , W_K|D)
 -\frac{1}{N}H(W_1,\dots , W_K|\mathbf{y}_E^{(1)},\dots , \mathbf{y}_E^{(K)},D)\\\nonumber
 \medmuskip=0mu
\thinmuskip=0mu
\thickmuskip=0mu
= \frac{1}{N}H(W_1,\dots , W_K)-\frac{1}{N}H(W_1,\dots , W_K|\mathbf{y}_E^{(1)},\dots , \mathbf{y}_E^{(K)},D)\\
\medmuskip=0mu
\thinmuskip=0mu
\thickmuskip=0mu
=\sm R_{\ell}-\frac{1}{N}H(W_1,\dots , W_K|\mathbf{y}_E^{(1)},\dots , \mathbf{y}_E^{(K)},D)\label{secondtermex}
\end{IEEEeqnarray}
Next, we bound the second term in (\ref{secondtermex}).
\begin{IEEEeqnarray}{l}
\nonumber\frac{1}{N}H(W_1,\dots , W_K|\mathbf{y}_E^{(1)},\dots , \mathbf{y}_E^{(K)},D)\\\nonumber
\medmuskip=0mu
\thinmuskip=0mu
\thickmuskip=0mu =\frac{1}{N}H(W_1,\dots , W_K,\mathbf{y}_E^{(1)},\dots , \mathbf{y}_E^{(K)}|D)
-\frac{1}{N}H(\mathbf{y}_E^{(1)},\dots , \mathbf{y}_E^{(K)}|D)\\\nonumber
\stackrel{(a)}{\geq}  \frac{1}{N}H(\bar{\mathbf{t}}_{1,2},\dots ,\bar{\mathbf{t}}_{\ell-1,\ell},\bar{\mathbf{t}}_{\ell+1,\ell},\dots, \bar{\mathbf{t}}_{K,K-1},\mathbf{y}_E^{(1)},\dots , \mathbf{y}_E^{(K)}|D)\\\nonumber -
\frac{1}{N}H(\mathbf{y}_E^{(1)},\dots , \mathbf{y}_E^{(K)}|D)\\\nonumber
\medmuskip=0mu
\thinmuskip=0mu
\thickmuskip=0mu
-\frac{1}{N}H(\bar{\mathbf{t}}_{1,2},\dots , \bar{\mathbf{t}}_{\ell-1,\ell},\bar{\mathbf{t}}_{\ell+1,\ell},\dots,\bar{\mathbf{t}}_{K,K-1}|\mathbf{y}_E^{(1)},\dots , \mathbf{y}_E^{(K)}, W_1,\dots , W_K,D)\\\nonumber
\medmuskip=0mu
\thinmuskip=0mu
\thickmuskip=0mu
\nonumber
\stackrel{(b)}{\geq}  \frac{1}{N}H(\bar{\mathbf{t}}_{1,2},\dots ,\bar{\mathbf{t}}_{\ell-1,\ell},\bar{\mathbf{t}}_{\ell+1,\ell},\dots, \bar{\mathbf{t}}_{K,K-1},\mathbf{y}_E^{(1)},\dots , \mathbf{y}_E^{(K)}|D)\\\nonumber-
\frac{1}{N}H(\mathbf{y}_E^{(1)},\dots , \mathbf{y}_E^{(K)}|D)- 2\epsilon_2
\\\nonumber
\medmuskip=0mu
\thinmuskip=0mu
\thickmuskip=0mu
\stackrel{(c)}{=} \frac{1}{N}H(\bar{\mathbf{t}}_{1,2},\dots , \bar{\mathbf{t}}_{K,K-1})-\frac{1}{N}H(\mathbf{y}_E^{(1)},\dots , \mathbf{y}_E^{(K)}|D)- 2\epsilon_2 \\
+\frac{1}{N}H(\mathbf{y}_E^{(1)},\dots , \mathbf{y}_E^{(K)}|D,\bar{\mathbf{t}}_{1,2},\dots, \bar{\mathbf{t}}_{K,K-1})\label{secterm}
\end{IEEEeqnarray}
Where inequality (a) comes from the chain rule. Note that in the sequence of $\bar{\mathbf{t}}_{1,2},\dots,\bar{\mathbf{t}}_{K,K-1}$, the indices $i=\ell$ are excluded. Soundness of inequality (b) is shown in Lemma~1 in Appendix. In short, it comes from applying Packing Lemma to the outer codewords which have a random i.i.d. structure. Equality (c) is due to chain rule.\\
\indent By substituting (\ref{secterm}) into (\ref{secondtermex}), we get
\begin{IEEEeqnarray}{L} 
\nonumber \frac{1}{N}I(W_1,\dots , W_K;\mathbf{y}_E|D) \leq \\\nonumber
\medmuskip=-1mu
\thinmuskip=-1mu
\thickmuskip=-1mu
\frac{1}{N}H\left(\bar{\mathbf{t}}_{1,1}, \dots , \bar{\mathbf{t}}_{K,K}\big|\mathbf{y}_E^{(1)},\dots,\mathbf{y}_E^{(K)},D\right)-\frac{1}{N}H(\bar{\mathbf{t}}_{1,2},\dots , \bar{\mathbf{t}}_{K,K-1})\\\nonumber
\medmuskip=-1mu
\thinmuskip=-1mu
\thickmuskip=-1mu
+\frac{1}{N}H(\mathbf{y}_E^{(1)},\dots , \mathbf{y}_E^{(K)}|D) 
-\frac{1}{N}H(\mathbf{y}_E^{(1)},\dots , \mathbf{y}_E^{(K)}|D,\bar{\mathbf{t}}_{1,2},\dots , \bar{\mathbf{t}}_{K,K-1})\\\nonumber
+2\epsilon_2+\epsilon_1\\\nonumber
\medmuskip=-1mu
\thinmuskip=-1mu
\thickmuskip=-1mu
=\frac{1}{N}H(\bar{\mathbf{t}}_{1,2},\dots , \bar{\mathbf{t}}_{K,K-1},\mathbf{y}_E^{(1)},\dots , \mathbf{y}_E^{(K)}|D)\\\nonumber
-\frac{1}{N}H(\bar{\mathbf{t}}_{1,2},\dots , \bar{\mathbf{t}}_{K,K-1},\mathbf{y}_E^{(1)},\dots , \mathbf{y}_E^{(K)}|D)+2\epsilon_2+\epsilon_1\\
=2\epsilon_2+\epsilon_1
\end{IEEEeqnarray}
In which $\epsilon_2$ and $\epsilon_1$ tend to zero for sufficiently large $N$. Therefore, the sum secure rate in (\ref{sumrate1}) provides weak secrecy; thus any sum secure rate satisfying
\begin{equation}
\sm R_{\ell}\leq\frac{1}{N}H\left(\bar{\mathbf{t}}_{1,1}, \dots , \bar{\mathbf{t}}_{K,K}\big|\mathbf{y}_E^{(1)},\dots,\mathbf{y}_E^{(K)},D\right)+\epsilon_1
\end{equation}
is also achievable with weak secrecy. Hence, we only need to show that the sum secure rate in (\ref{th2}) is a lower bound on (\ref{sumrate1}). We have
\begin{IEEEeqnarray}{l}
\nonumber \frac{1}{N}H\left(\bar{\mathbf{t}}_{1,1}, \dots , \bar{\mathbf{t}}_{K,K}\big|\mathbf{y}_E^{(1)},\dots,\mathbf{y}_E^{(K)},D\right)+\epsilon_1\\
\medmuskip=0mu
\thinmuskip=0mu
\thickmuskip=0mu
\nonumber \geq \frac{1}{N}H\left(\bar{\mathbf{t}}_{1,1}, \dots , \bar{\mathbf{t}}_{K,K}\right)-\frac{1}{N}H\left(\mathbf{y}_E^{(1)},\dots,\mathbf{y}_E^{(K)}\big|D\right)+\epsilon_1\\
\nonumber
=\frac{1}{N}H\left(\bar{\mathbf{t}}_{1,2}, \dots , \bar{\mathbf{t}}_{K,K-1}\right)+\frac{1}{N}\sm H\left(\bar{\mathbf{t}}_{\ell,\ell}\right)
\\
\nonumber
-\frac{1}{N}H\left(\mathbf{y}_E^{(1)},\dots,\mathbf{y}_E^{(K)}\big|D\right)+\epsilon_1 \\
\medmuskip=0mu
\thinmuskip=0mu
\thickmuskip=0mu
= \sum_{k=2}^{K(K-1)+1}R_{comb,k}+\frac{1}{N}\sm H\left(\bar{\mathbf{t}}_{\ell,\ell}\right)\label{beforetermye}\\
-\frac{1}{N}H\left(\mathbf{y}_E^{(1)},\dots,\mathbf{y}_E^{(K)}\big|D\right)+\epsilon_1 \label{termye}
\end{IEEEeqnarray}
Next, we bound the first term in (\ref{termye}). We have
\begin{IEEEeqnarray}{l}
\nonumber \frac{1}{N}H(\mathbf{y}_E^{(1)},\dots , \mathbf{y}_E^{(K)}\big|D) \leq \sum_{\ell=1}^K\frac{1}{N}H(\mathbf{y}_E^{(\ell)}|D)\\
\nonumber
=\frac{1}{N}\sum_{\ell=1}^K H\left(\left[\mathbf{y}_E^{(\ell)}\right]\modnll, Q_{\Lambda_{(\ell,\ell)}}(\mathbf{y}_E^{(\ell)})\big|D \right)
\\
\nonumber
\leq \frac{1}{N}\sum_{\ell=1}^K H \left(\left[\mathbf{y}_E^{(\ell)}\right]\modnll \big |\bar{\mathbf{d}}_{\ell,\ell},D\backslash\{\bar{\mathbf{d}}_{\ell,\ell}\}\right)\\
\nonumber 
+\frac{1}{N} \sum_{\ell=1}^K H \left( Q_{\Lambda_{(\ell,\ell)}}(\mathbf{y}_E^{(\ell)})\right)
\end{IEEEeqnarray}
\begin{IEEEeqnarray}{l}
\nonumber
\medmuskip=-1.5mu
\thinmuskip=-1.5mu
\thickmuskip=-1.5mu
= \tiny{\frac{1}{N}\sum_{\ell=1}^K H \left(\left[\bar{\mathbf{t}}_{\ell,\ell}+\sum_{j \neq \ell}\left[\bar{\mathbf{t}}_{j,\ell}+\bar{\mathbf{d}}_{j,\ell}\right] \modnjl \right]\modnll \Bigg |D\right)}\\
\nonumber
+\frac{1}{N} \sum_{\ell=1}^K H \left( Q_{\Lambda_{(\ell,\ell)}}(\mathbf{y}_E^{(\ell)})\right)
\\
\nonumber
\medmuskip=-2mu
\thinmuskip=-2mu
\thickmuskip=-2mu
=\frac{1}{N}\sum_{\ell=1}^K H \left(\left[\bar{\mathbf{t}}_{\ell,\ell}+\sum_{j \neq \ell}\bar{\mathbf{t}}_{j,\ell}-\sum_{j \neq \ell} Q_{\Lambda_{(j,\ell)}}(\bar{\mathbf{t}}_{j,\ell}+\bar{\mathbf{d}}_{j,\ell})\right]\modnll \Bigg |D \right)\\
\nonumber
+\frac{1}{N} \sum_{\ell=1}^K H \left( Q_{\Lambda_{(\ell,\ell)}}(\mathbf{y}_E^{(\ell)})\right)\\
\nonumber
\medmuskip=0mu
\thinmuskip=0mu
\thickmuskip=0mu
\stackrel{(d)}{=} \frac{1}{N}\sum_{\ell=1}^K H \left(\left[\bar{\mathbf{t}}_{\ell,\ell}+\bar{\mathbf{q}}_{\ell} \right]\modnll \right)
+\frac{1}{N} \sum_{\ell=1}^K H \left( Q_{\Lambda_{(\ell,\ell)}}(\mathbf{y}_E^{(\ell)})\right)\\\nonumber
\stackrel{(e)}{=} \frac{1}{N}\sum_{\ell=1}^K H \left(\left[\bar{\mathbf{t}}_{\ell,\ell} \right]\modnll \right)
+\frac{1}{N} \sum_{\ell=1}^K H \left( Q_{\Lambda_{(\ell,\ell)}}(\mathbf{y}_E^{(\ell)})\right)\\\nonumber
=\frac{1}{N}\sum_{\ell=1}^K H(\bar{\mathbf{t}}_{\ell,\ell})+\frac{1}{N} \sum_{\ell=1}^K H \left( Q_{\Lambda_{(\ell,\ell)}}(\mathbf{y}_E^{(\ell)})\right)\\
\nonumber
\stackrel{(f)}{\leq} \frac{1}{N}\sum_{\ell=1}^K H(\bar{\mathbf{t}}_{\ell,\ell})\\
\medmuskip=0mu
\thinmuskip=0mu
\thickmuskip=0mu
+\frac{1}{2} \sum_{\ell=1}^K \log \left(\frac{\sum_{j=1, j \neq \ell}^{K} \gamma^2_{j,\ell}P_{j,\ell}+\gamma^2_{\ell,\ell}P_{\ell,\ell}}{\gamma^2_{\ell,\ell}P_{\ell,\ell}}\right)
+\delta(\epsilon) \label{finaleq}
\end{IEEEeqnarray}
Where in equality (d), the random vector $\bar{\mathbf{q}}_{\ell}$ has been defined as
\begin{equation}
\bar{\mathbf{q}}_{\ell} \triangleq \left[\sum_{j \neq \ell}\bar{\mathbf{t}}_{j,\ell}-\sum_{j \neq \ell} Q_{\Lambda_{(j,\ell)}}(\bar{\mathbf{t}}_{j,\ell}+\bar{\mathbf{d}}_{j,\ell})\right]\modnll
\end{equation} 
Furthermore, equality (e) comes from conditions in (\ref{condition1}) along with the Crypto Lemma [Lemma 2,~\cite{forney2003role}] in which the compact group satisfying in the lemma's conditions is the set $\mathcal{L}_{\ell,\ell}= \Lambda_{f(,\ell,\ell)} \cap \mathcal{V}_{(\ell,\ell)}$. Lastly, inequality (f) is deduced from Lemma~1 in~\cite{paper1}. Here $\gamma_{j,i}$ is defined as in (\ref{align1new}) and (\ref{align2new}).

\indent Consequently, expressions (\ref{beforetermye}), (\ref{termye}), and (\ref{finaleq}) yield
\begin{IEEEeqnarray}{l}
\nonumber\frac{1}{N}H\left(\bar{\mathbf{t}}_{1,1}, \dots , \bar{\mathbf{t}}_{K,K}\big|\mathbf{y}_E^{(1)},\dots,\mathbf{y}_E^{(K)},D\right)+\epsilon_1\\\nonumber
\geq \sum_{k=2}^{K(K-1)+1}R_{comb,k}\\\nonumber
-\frac{1}{2} \sum_{\ell=1}^K \log \left(\frac{\sum_{j=1, j \neq \ell}^{K} \gamma^2_{j,\ell}P_{j,\ell}+\gamma^2_{\ell,\ell}P_{\ell,\ell}}{\gamma^2_{\ell,\ell}P_{\ell,\ell}}\right)
-\delta(\epsilon)+\epsilon_1
\end{IEEEeqnarray}
In which $\epsilon_1$ and $\delta(\epsilon)$ tend to zero for sufficiently large $N$. This completes the proof of weak secrecy of Theorem~1. \hfill $\blacksquare$ 
\section{Simulation results}\label{sec5}
In this section, we evaluate the performance of our proposed scheme numerically. To this end, we compute our achievable results with and without security for a three-user Gaussian wiretap MAC and then compare the achievable sum secure rate with the result implied by i.i.d. Gaussian random codes in~\cite{tekin2008general} as well as the result in~\cite{paper1}. Furthermore, we show experimentally that our achievable sum rate, i.e., $\sum_{k=1}^{K(K-1)+1}R_{comb,k}$ reaches Gaussian MAC capacity in high-SNR regimes. \\
\indent In the first experiment,  we compare our sum secure rate with the one achieved by i.i.d. Gaussian random codes. To this end, we have considered a Three-user GW-MAC in which the channel gain vectors, $\mathbf{h}$, $\mathbf{g}$, were chosen as realizations of i.i.d. normal distributions. The simulation was run for $1000$ instances and the average sum rates are shown as a function of SNR in Figure 3. \\
\begin{figure}
\vspace{-4mm}
\centering
\includegraphics[height= 1.6 in, width=0.5\textwidth]{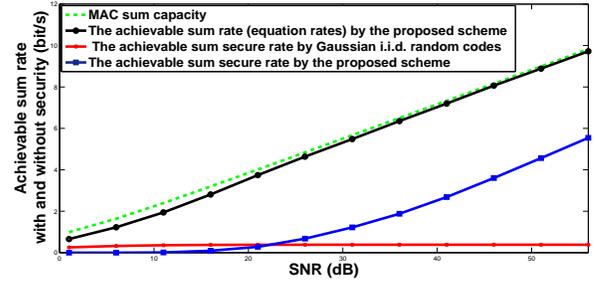}
\vspace{-7mm}
\centering
\caption{\small {Achievable sum rate, with and without security evaluated for a three-user Gaussian wiretap MAC at different SNR.}}
\label{fig3}
\vspace{-2mm}
\end{figure}
\indent For simplicity, in implementing the achievable result of our scheme, we allocated powers among sub-message codewords equally, i.e., for $\ell \in \{1,\dots,K\}$ we set $P_{\ell,i}=\frac{P}{K},\forall~i$.  Due to this simplification, the sum secure rate displayed as our achievable result in Figure~3 and Figure~4 are lower bounds on the highest sum secure rate that can be achieved by (\ref{th2}). Also, similar to~\cite{ordentlich2014approximate}, we have approximated the best integer coefficient vectors $\mathbf{a}^{(k)}$ for all equations using the LLL reduction algorithm which forms a set of $K(K-1)+1$ linearly independent lattice vectors~\cite{LenstraLenstraLovasz1982}. \\
\indent A comparison between the curve displayed as our achievable sum secure rate and the one related to the achievable sum secure rate offered by i.i.d. Gaussian random codes reveals the advantage of our scheme versus pure random coding in moderate and high SNR regimes. Moreover, to shed light on the performance of our decoding strategy, we plotted the sum of the integer combinations rates, i.e., $\sum_kR_{comb,k}$ as well. As it can be seen, our achievable sum rate (without security) reaches Gaussian MAC capacity in the high SNR regimes. Recall that the compute-and-forward decoding strategy in~\cite{ordentlich2014approximate} offers a lower bound on sum rate that achieves MAC sum capacity within a constant gap. Figure~3 shows numerically that our extension to their decoding strategy yields the same performance in terms of the achievable sum rate without security. \\
\indent Also, Figure~4 illustrates the advantage of our new scheme over the one in~\cite{paper1}. The improvement can be clearly seen in high SNR regimes. This result is expected as the sum secure degrees of freedom achieved by our new scheme is higher than the one achieved by~\cite{paper1}. Note that due to the sub-optimal power allocation considered in the numerical implementation, the red curve displayed in Figure~4 is a lower bound on the maximum sum secure rate that can be achieved by our proposed scheme. Therefore, given the fact that the average slope of the red curve is much bigger than for the blue curve, our achievable sum secure rate under the optimal power allocation crosses the blue curve at lower SNR values. Furthermore, the location of the crossing point is also dependent on the number of users.
\begin{figure}
\vspace{-4mm}
\centering
\includegraphics[height=1.6 in, width=0.5\textwidth]{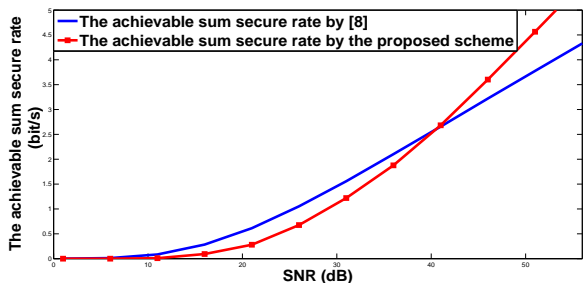}
\vspace{-7mm}
\centering
\caption{\small {Achievable sum secure rate for a three-user Gaussian wiretap MAC at different SNR.}}
\label{fig4}
\vspace{-2mm}
\end{figure}
\section{Conclusion}\label{sec6}
In this work, we considered the $K$-user Gaussian wiretap multiple-access channel. We developed a new scheme based on an extension of the compute-and-forward decoding strategy, nested lattice structure, and cooperative jamming. Our scheme consists of two layers: the inner layer which is the nested lattice coding structure and the outer layer which is the random coding. In short, our scheme achieves weak secrecy by means of three factors including proper lattice alignment, cooperative jamming signals, and i.i.d. repetitions of lattice codebook. Note that the latter has been added only to prove weak secrecy for the finite-SNR regime. Furthermore, we derived a new lower bound on sum secure rate for the finite-SNR regime based on our scheme and showed that it achieves the optimal sum secure degrees of freedom.
\section*{Acknowledgment}
The authors would like to thank Bobak Nazer and Prakash Ishwar for their valuable comments and helpful discussions.
\bibliography{refn}
\section{Appendix}\label{sec7}
\subsection{Proof of Corollary~1}\label{sec7.1}
We begin the proof by showing that $\sum_{k=1}^{K(K-1)+1}R_{comb,k}$ scales with $\log(P)$. We have
\begin{IEEEeqnarray}{l}
\nonumber \sum_{k=1}^{K(K-1)+1}R_{comb,k}=\sum_{k=1}^{K(K-1)+1}\frac{1}{2}\log\left(\frac{\Gamma(k,k)^2P}{\|\mathbf{F}.\mathbf{a}^{(k)}\|^2}\right)\\\nonumber
=\frac{K(K-1)+1}{2}\log(P)+\frac{1}{2}\log\left(\prod_k \Gamma(k,k)^2\right)\\\nonumber
-\frac{1}{2}\log\left(\prod_k \|\mathbf{F}.\mathbf{a}^{(k)}\|^2\right)\\\nonumber
\stackrel{(a)}{\geq}\frac{K(K-1)+1}{2}\log(P)+\frac{1}{2}\log\left(\prod_k \Gamma(k,k)^2\right)\\\nonumber
-\frac{1}{2}\log\left((K(K-1)+1)^{(K(K-1)+1)}.|\det(\mathbf{F})|^2\right)\\\nonumber
=\frac{K(K-1)+1}{2}\log(P)+\frac{1}{2}\log\left(\prod_k \Gamma(k,k)^2\right)\\
\nonumber
-(K(K-1)+1)\frac{1}{2}\log\left(K(K-1)+1\right)\\\nonumber
-\frac{1}{2}\log\left(\prod_k \Gamma(k,k)^2\right)-\frac{1}{2}\log\left(\det\left(\frac{1}{P}\Gamma^{-2}+\tilde{\mathbf{h}}\tilde{\mathbf{h}}^T\right)\right)\\\nonumber
\medmuskip=0mu
\thinmuskip=0mu
\thickmuskip=0mu
=\frac{K(K-1)+1}{2}\log(P)-(K(K-1)+1)\frac{1}{2}\log\left(K(K-1)+1\right)\\
\nonumber
-\frac{1}{2}\log\left(\det\left(\Gamma^{-1}(\frac{1}{P}\mathbf{I}+\Gamma\tilde{\mathbf{h}}\tilde{\mathbf{h}}^T\Gamma)\Gamma^{-1}\right)\right)\\\nonumber
\medmuskip=0mu
\thinmuskip=0mu
\thickmuskip=0mu
=\frac{K(K-1)+1}{2}\log(P)-(K(K-1)+1)\frac{1}{2}\log\left(K(K-1)+1\right)\\
\nonumber
-\frac{1}{2}\log\left(\det\left(\Gamma^{-2}\right)\right)-\frac{1}{2}\log\left(\det\left(\frac{1}{P}\mathbf{I}+\Gamma\tilde{\mathbf{h}}\tilde{\mathbf{h}}^T\Gamma\right)\right)\\\nonumber
\stackrel{(b)}{=}\frac{K(K-1)+1}{2}\log(P)-\frac{1}{2}\log\left(\frac{P^{K(K-1)+1}}{1+\|\Gamma.\mathbf{\tilde{h}}\|^2P}\right)\\\nonumber
-(K(K-1)+1)\frac{1}{2}\log\left(K(K-1)+1\right)-\frac{1}{2}\log\left(\det\left(\Gamma^{-2}\right)\right)\\
\nonumber
\medmuskip=0mu
\thinmuskip=0mu
\thickmuskip=0mu
=\frac{1}{2}\log\left(1+\|\Gamma.\mathbf{\tilde{h}}\|^2P\right)-\frac{1}{2}\log\left(\det\left(\Gamma^{-2}\right)\right)\\
-(K(K-1)+1)\frac{1}{2}\log\left(K(K-1)+1\right)\label{38}
\end{IEEEeqnarray}
In the above arguments, inequality (a) is concluded from Minkowski Theorem in~\cite{ordentlich2014approximate} and equality (b) is the result of applying Sylvester's determinant identity (refer e.g. to \cite{harville1997matrix}).\\
\indent Note that from (\ref{38}) two points can be deduced: First, our achievable sum rate (without security) is within a constant gap (with respect to $P$) from $K$-user MAC sum capacity\footnote{To see this clearly, choose $\gamma_{\ell,i}=1$ and $P_{\ell,i}=\frac{P}{K}$, for all $i,\ell$.} and second, $\medmuskip=0mu
\thinmuskip=0mu
\thickmuskip=0mu \sum_{k=1}^{K(K-1)+1}R_{comb,k} \propto \log(P)$. Now, using Corollary~5 in~\cite{ordentlich2014approximate}, for large enough $P$, we have 
\begin{equation}\label{singlercomb}
R_{comb,k}\leq \frac{1}{K(K-1)+1}\log(P)~~\forall k
\end{equation}
\indent As a result, (\ref{38}) and (\ref{singlercomb}) together yield $\sum_{k=2}^{K(K-1)+1}R_{comb,k} \propto \log(P)$. This completes the proof of Corollary~1. \hfill $\blacksquare$
\subsection{Proof of Corollary~2}\label{sec7.2}
We derive the sum secure degrees of freedom in two steps. In step one, we show that the limit of the first term in~(\ref{th2}) when $P \rightarrow \infty$ is $\frac{K(K-1)}{K(K-1)+1}$. As it was mentioned earlier, the jamming codewords get aligned at the receiver's side and as a result, the receiver decodes $K(K-1)+1$ equations. This situation is equivalent to the multiple-access channel with $K(K-1)+1$ users in which user one has the highest rate of $R_{comb,1}$ and other users operate at rates equal to $R_{comb,k}$ for $\{2,\dots,K(K-1)+1\}$. Thus, similar to the arguments in the proof of Corollary~1 and by using Corollary~5 in~\cite{ordentlich2014approximate}, we have
\begin{equation}
\limsup_{P\rightarrow \infty}\frac{R_{comb,k}}{\frac{1}{2}\log(1+P)}= \frac{1}{K(K-1)+1},~~\forall~k
\end{equation}
Therefore,
\vspace{- 2 mm}
\begin{equation}
\medmuskip=0mu
\thinmuskip=0mu
\thickmuskip=0mu
\limsup_{P\rightarrow \infty}\left(\sum_{k=2}^{K(K-1)+1} R_{comb,k}\right)=\frac{K(K-1)}{K(K-1)+1}
\end{equation}
Next, we show that the second term in~(\ref{th2}) is a constant with respect to the power $P$ and hence it does not contribute to the sum secure degrees of freedom. To this end, let us consider a sub-optimal power allocation strategy in which user~$\ell$ allocates its total power equally among its sub-codewords, i.e., $P_{\ell,i}=\frac{P}{K},~\forall~i\in\{1,\dots,K\}$. Therefore, we have
\begin{equation}
\frac{\sum_{j=1,  j \neq \ell}^{K} \gamma^2_{(j,\ell)}P_{(j,\ell)}+\gamma^2_{(\ell,\ell)}P_{\ell,\ell}}{\gamma^2_{(\ell,\ell)}P_{\ell,\ell}}=\sum_{j=1 \atop j \neq \ell}^{K} \frac{\gamma^2_{(j,\ell)}+\gamma^2_{(\ell,\ell)}}{\gamma^2_{(\ell,\ell)}}
\end{equation}
As a result, the second term in~(\ref{th2}) is reduced to $\frac{1}{2}\sm\log\left(\sum_{j=1 \atop j \neq \ell}^{K} \frac{\gamma^2_{(j,\ell)}+\gamma^2_{(\ell,\ell)}}{\gamma^2_{(\ell,\ell)}}\right)$, which is a constant with respect to $P$, hence, the proof of Corollary~2 is completed.$\blacksquare$
\subsection{Supplementary lemma}\label{sec7.3}
\begin{lemma}
For the achievable scheme presented in Section VII, we have 
 \begin{equation}
 \frac{1}{nB}H(\bar{\mathbf{t}}_{1,2},\dots,\bar{\mathbf{t}}_{K,K-1}|W_1,\dots,W_K,\mathbf{y}_E,D)\leq 2\epsilon_2, \label{ps}
 \end{equation}
 where $\epsilon_2$ goes to zero if $B$ is taken large enough.
 \end{lemma}
\indent \textit{Proof:} We prove Lemma~1 by showing that 
\vspace{-2 mm}
 \begin{equation}\label{endbeforej}
 \frac{1}{nB}H(\bar{\mathbf{t}}_{1,1},\bar{\mathbf{t}}_{1,2},\dots,\bar{\mathbf{t}}_{K,K}|W_1,\dots,W_K,\mathbf{y}_E,D)\leq 2\epsilon_2
 \end{equation}
Then, the correctness of Lemma~1 is automatically deduced.\\
\indent Let us assume that for all $(\ell,i) \in \{1,\dots,K\} \times \{1,\dots,K\}$ each codeword $\bar{\mathbf{t}}_{\ell,i}$ is uniquely identified with two indices $(w_{\ell,i},w^{\prime}_{\ell,i})$. The  corresponding index variable for $w_{\ell,i}$ is $W_{\ell,i},~\forall i$. Also, assume that $H(w_{\ell,\ell})=0,~\forall \ell$ and $H(\{w_{\ell,i}\}_{i=1, i \neq \ell}^K)=H(W_{\ell})$. Assume that $1\leq w_{\ell,i}\leq 2^{NR_{\ell,i}}$ and $1\leq w^{\prime}_{\ell,i}\leq 2^{NR^{\prime}_{\ell,i}}$, where $N\sum_{(\ell,i)} R_{\ell,i}=N\sum_{\ell}R_{\ell}=H\left(\bar{\mathbf{t}}_{1,1}, \dots , \bar{\mathbf{t}}_{K,K}\big|\mathbf{y}_E^{(1)},\dots,\mathbf{y}_E^{(K)},D\right)+N\epsilon_1$ and $N\sum_{(\ell,i)}R^{\prime}_{\ell,i}=I\left(\bar{\mathbf{t}}_{1,1}, \dots , \bar{\mathbf{t}}_{K,K};\mathbf{y}_E^{(1)},\dots,\mathbf{y}_E^{(K)}\big|D\right)-N\epsilon_1$, where $\epsilon_1>0$. Note that $N\sum_{(\ell,i)} (R^{\prime}_{\ell,i}+R_{\ell,i})=H(\bar{\mathbf{t}}_{1,1}, \dots , \bar{\mathbf{t}}_{K,K})=N\sum_{k=1}^{K(K-1)+1}R_{comb,k}$. \\
\indent Having the bin indices $(w_1,\dots ,w_K)$, the eavesdropper needs to look for the transmitted codewords in the corresponding sub-codebooks $\left(\mathcal{C}_1(w_1),\dots,\mathcal{C}_K(w_K)\right)$. From the above setting, the number of codewords $\bar{\mathbf{t}}_{\ell,i}$ for the eavesdropper to check would be $2^{B\left(I\left(\mathbf{t}_{1,1}, \dots , \mathbf{t}_{K,K};\mathbf{y}_{E,<n>}^{(1)},\dots,\mathbf{y}_{E,<n>}^{(K)}\big|\mathbf{d}_{1,1}, \dots , \mathbf{d}_{K,K}\right)-n\epsilon_1\right)}$, where the subscript $\medmuskip=0mu
\thinmuskip=0mu
\thickmuskip=0mu <n>$ denotes an $n$-length block of the corresponding random vector. Among these remaining codewords, the eavesdropper looks for those ones that satisfy in the following condition.
\begin{IEEEeqnarray*}{l}
\medmuskip=-1mu
\thinmuskip=-1mu
\thickmuskip=-1mu
\nonumber \left(\bar{\mathbf{t}}_{1,1}, \dots , \bar{\mathbf{t}}_{K,K},\mathbf{y}_E^{(1)},\dots,\mathbf{y}_E^{(K)},D\right)
\in \mathcal{T}_{\epsilon_2}^B\left(P_{\bar{\mathbf{t}}_{1,1}, \dots , \bar{\mathbf{t}}_{K,K},\mathbf{y}_E^{(1)},\dots,\mathbf{y}_E^{(K)}|D}\right)
\end{IEEEeqnarray*}
where $\mathcal{T}_{\epsilon_2}^B(P_{\bar{\mathbf{t}}_{1,1}, \dots , \bar{\mathbf{t}}_{K,K},\mathbf{y}_E^{(1)},\dots,\mathbf{y}_E^{(K)}|D})$ is the set of $\epsilon_2$-jointly typical sequences. Without loss of generality, let us assume that $\bar{\mathbf{t}}_{1,1}=\bar{\mathbf{t}}_{1,1}^*,...,\bar{\mathbf{t}}_{K,K}=\bar{\mathbf{t}}_{K,K}^*$ are sent. For ease of notation, define ${\bar{\mathbf{t^*}}_1}^K\triangleq (\bar{\mathbf{t}}_{1,1}^*,\dots ,\bar{\mathbf{t}}_{K,K}^*)$. Then, a decoding error would occur in either of the following two possible events.
\begin{IEEEeqnarray*}{l}
\medmuskip=-1mu
\thinmuskip=-1mu
\thickmuskip=-1mu
\nonumber \mathcal{E}_1=\bigg\lbrace\left({\bar{\mathbf{t^*}}_1}^K,\mathbf{y}_E,\mathbf{d}_1^K\right) \not \in \mathcal{T}_{\epsilon_2}^B(P_{{\bar{\mathbf{t}}_1}^K,\mathbf{y}_E|\mathbf{d}_1^K}) \bigg\rbrace\\
\begin{split}
\mathcal{E}_2= \bigg\lbrace
 \nonumber \exists\left({\bar{\mathbf{t}}_1}^K,\mathbf{y}_E,D\right) \in \mathcal{T}_{\epsilon_2}^B(P_{{\bar{\mathbf{t}}_1}^K,\mathbf{y}_E|D}): \\
 ~{\bar{\mathbf{t}}_1}^K \neq {\bar{\mathbf{t^*}}_1}^K,\mathbf{t}_{\ell} \in \mathcal{C}_{\ell}(w_\ell),\ell \in \{1, \dots ,K\} \bigg\rbrace
\end{split}
\end{IEEEeqnarray*}
 By the AEP theorem, the first error event is bounded above by $\epsilon_2$, and the second term can also be bounded by applying the Packing Lemma, [lemma 3.1,~\cite{el2011network}] to codewords $\bar{\mathbf{t}}_{1,1}, \dots ,\bar{\mathbf{t}}_{K,K}$, i.e.,
\begin{IEEEeqnarray}{l}
\nonumber \medmuskip=0mu
\thinmuskip=0mu
\thickmuskip=0mu \mathbb{P}\left\lbrace \mathcal{E}_2 \right\rbrace 
\leq  2^{B\left(I\left(\mathbf{t}_{1,1}, \dots , \mathbf{t}_{K,K};\mathbf{y}_{E,<n>}^{(1)},\dots,\mathbf{y}_{E,<n>}^{(K)}\big|\mathbf{d}_{1,1}, \dots , \mathbf{d}_{K,K}\right)-n\epsilon_1\right)} \\
\nonumber
\medmuskip=0mu
\thinmuskip=0mu
\thickmuskip=0mu
 \times 2^{-B\left(I\left(\mathbf{t}_{1,1}, \dots , \mathbf{t}_{K,K};\mathbf{y}_{E,<n>}^{(1)},\dots,\mathbf{y}_{E,<n>}^{(K)}\big|\mathbf{d}_{1,1}, \dots , \mathbf{d}_{K,K}\right)-\delta(\epsilon_2)\right)}\\\nonumber
\leq 2^{-B\big( n\epsilon_1-\delta(\epsilon_2)\big)},
\end{IEEEeqnarray} 
where $\delta(\epsilon_2)$ tends to zero as $\epsilon_2$ goes to zero. Now, choose $n \epsilon_1 \geq \delta(\epsilon_2)$. Hence, for sufficiently large $B$, the total probability of the error events will be upper-bounded as $\mathrm{P}_e \leq 2\epsilon_2$. As a result, using Fano's inequality in~\cite{el2011network}, the correctness of (\ref{endbeforej}) is concluded.
\hfill $\blacksquare$
\end{document}